\documentclass[conference]{IEEEtran}

\IEEEoverridecommandlockouts

\usepackage{graphicx}
\usepackage{amsmath}
\usepackage{amssymb}
\usepackage{color}
\usepackage{setspace}
\usepackage{bbm}
\usepackage{bm}
\usepackage{algorithm}
\usepackage{algpseudocode}

\usepackage{amsmath}

\usepackage{svg}

\usepackage{cite}
\usepackage{url}

\usepackage{algorithm}
\usepackage{algpseudocode}
\algnewcommand\algorithmicinput{\textbf{Input:}}
\algnewcommand\INPUT{\item[\algorithmicinput]}
\algnewcommand\algorithmicoutput{\textbf{Output:}}
\algnewcommand\OUTPUT{\item[\algorithmicoutput]}

\usepackage{mathtools}

\graphicspath{{figures/}}
\usepackage{subcaption}

\def\BibTeX{{\rm B\kern-.05em{\sc i\kern-.025em b}\kern-.08em
    T\kern-.1667em\lower.7ex\hbox{E}\kern-.125emX}}

\usepackage[left=1.5cm,right=1.5cm,
  top=1.5cm,bottom=1.5cm,bindingoffset=0cm]{geometry}
\begin{document}


\title{ Trainable Least Squares to Reduce PAPR \\ in OFDM-based Hybrid Beamforming Systems
\thanks{The research was carried out at Skoltech and supported by the Russian Science Foundation (project no. 24-29-00189).
}}

\author{Andrey~Ivanov,
        Alexander~Osinsky, 
        Roman~Bychkov,
        Vladimir~Kalinin,
        and Dmitry~Lakontsev
        \thanks{A. Ivanov,
        A. Osinsky, 
        R. Bychkov,
        V. Kalinin,
        and D. Lakontsev are with Skolkovo Institute of Science and Technology, Moscow, Russia. E-mails: an.ivanov@skoltech.ru,
        alexander.osinsky@skoltech.ru, R.Bychkov@skoltech.ru,  
        kkalinin@list.ru,
        d.lakontsev@skoltech.ru}
        }

\maketitle
\IEEEdisplaynontitleabstractindextext

\begin{abstract}
In this paper, we propose a trainable least squares (LS) approach for reducing the peak-to-average power ratio (PAPR) of orthogonal frequency division multiplexing (OFDM) signals in a hybrid beamforming (HBF) system. 

Compared to digital beamforming (DBF), in HBF technology the number of antennas exceeds the number of digital ports. Therefore, PAPR reduction capabilities are restricted by both a limited bandwidth and the reduced size of digital space. The problem is to meet both conditions. Moreover, the major HBF advantage is a reduced system complexity, thus the complexity of the PAPR reduction algorithm is expected to be low.

To justify the performance of the proposed trainable LS, we provide a performance bound achieved by convex optimization using the CVX Matlab package. Moreover, the complexity of the proposed algorithm can be comparable to the minimal complexity of the digital ``twin'' calculation in HBF. The abovementioned features prove the feasibility of the trained LS for PAPR reduction in fully-connected HBF.

\end{abstract}
 \vskip 0.2cm


\section{Introduction}

\IEEEPARstart{O}{rthogonal} frequency division multiplexing (OFDM) is a well-known information transmission technique and a basis of numerous wireless communications systems such as 5G, Wi-Fi, and others \cite{5G}. It is widely employed in the downlink channel of modern wireless communications. Unfortunately, OFDM signals suffer from a high peak-to-average power ratio (PAPR), defined as the ratio of peak signal sample power to mean power among all samples in an OFDM symbol:
\begin{equation}
PAPR [\bm x(n)] =\frac{\max({|\bm x(n)|}^2)}{\mathbb{E}({|\bm x(n)|}^2)},
\end{equation}
where $\bm x(n) $ is the time domain OFDM symbol, represented by $N_{FFT}$ samples, $\ n \in \left[1\dots N_{FFT}\right]$ is the sample index of the time domain OFDM symbol, $\max \left(\cdot\right)$ is the maximum value operation, $\mathbb{E}\left(\cdot\right)$ is the expectation operator.

High PAPR causes high power amplifiers (HPAs) to operate nonlinearly, which saturates the amplifier at large input signal amplitudes \cite{hpa}. Consequently, the HPA produces an additive non-linear noise. Thus, the HPA for the high PAPR problem needs to either have a large linear range or employ a successful PAPR reduction algorithm. Furthermore, using digital-to-analog converters (DAC) with better resolution on the transmitter (TX) end is required when the PAPR value is large \cite{DAC_ADC_PAPR}.

Fast-growing complexity and power consumption become an obstacle on the way to multi-antenna transmitters \cite{Molisch, Lyashev}. To overcome this problem, a hybrid beamforming (HBF) scheme stands as the best candidate for precoding \cite{Sohrabi}. It employs both digital and analog beamforming \cite{yaliny, phy_hbf}. The second is implemented via phase-shifters that reduce an overall complexity \cite{Ahmed}. 

Let us note, that compared to the fully digital precoding, in HBF the antenna signal is not presented in the digital domain which complicates the PAPR reduction. 

\subsection{PAPR reduction methods}

There is very limited research on PAPR reduction in HBF. Thus, we describe cutting-edge PAPR reduction techniques that are 5G and 6G compatible.

The most popular and simple method for reducing PAPR is clipping and filtering (C$\&$F). There are two distinct kinds of signal distortions produced by clipping because of the non-linear nature of the technique: in-band distortions and out-of-band radiation \cite{CF1, CF2}. Filtration reduces out-of-band radiation. Sadly, it also results in fresh growth peaks.

A PAPR compensation signal is constructed from a sequence of reserved subcarriers using the tone reservation (TR) technique, which uses an iterative amplitudes computation \cite{PW1}. Nevertheless, \cite{TR_FISTA}, computational complexity is unreasonably high. The non-iterative selective tone reservation (STR) approach based on minimal mean squared error (MMSE) is unaffected by this problem. Furthermore, the efficiency of this technique is about at maximum \cite{STR}.

There have been initiatives \cite{TR_DL, TR_DL1} to use deep learning techniques to enhance TR. This algorithm unfolds each iteration of the classical TR strategy into a layer of a deep neural network (deep NN, DNN). Almost always, the clipping threshold is chosen at each iteration. Despite having the same computing complexity and amount of iterations as normal TR, this approach outperforms it in terms of PAPR reduction capabilities.

In a different class of approaches, the PAPR reduction signal is generated using free spatial directions \cite{PAPR_MIMO_DL}. Extra benefit in the downlink of MU-MIMO applications is provided by the \cite{UBR} method for unused beam reserve (UBR), which makes use of free spatial beams. By employing non-iterative amplitude modulation of idle beams without users, it is possible to avoid signal deterioration for target users while lowering PAPR for every transmitting antenna. UBR is also compatible with 5G and provides reduced computational latency. 

Initiatives are now underway to support the existing PAPR reduction algorithms using machine learning (ML) techniques. Such methods, \cite{IVANOV2,krikunov} complements the STR and UBR methodologies. They replace the exhaustive search for the optimal hyperparameters approximation in the low-parameter model with standard machine learning algorithms that have low computational complexity. Finding the ideal hyperparameters is quick, and the model might be applied to the 5G communications infrastructure.

\subsection{Our contribution}

The problem of PAPR reduction in HBF is using limited bandwidth and a limited subspace to approximate sparse peaks existing in full-band in full-space. 

To overcome this problem, we divide it into 2 consequent steps. Firstly, we examine the band-limited generation of PAPR reduction vectors for each antenna. Then we construct a matrix of these signals and fit the achieved band-limited PAPR reduction matrix into a limited subspace using trainable least squares (LS) transformation.

For training, a Genetic Algorithm (GA) was employed. To validate performance, the optimal bound was also calculated. Complexity analysis confirms the algorithm feasibility for future generation systems.  

\section{Single antenna -- tone reservation} \label{sec:model}
Let's analyze the STR algorithm \cite{STR} as a candidate for HBF. It first finds a signal consisting of only peaks for further reduction. To select these peaks, the thresholding is applied to absolute values of the antenna signal ${\bm{x}} \in \mathbb{C}^{N_{FFT}}$ as follows: 
\begin{align}
{y}(n) &= \begin{cases} x(n) \left( 1 - \frac{\tau}{\left| x(n) \right|} \right),& \text{if } |{x}(n)| \geq \tau, \\ 0, &\text{otherwise}, \end{cases} \label{STR1}
\end{align}
where $\bm{x}(n)$ is the $n$'th element of $\bm x$, $\tau$ is the threshold to optimize,
${\bm{y}} \in \mathbb{C}^{N_{FFT}}$, consisting of undesired peaks of ${\bm{x}}$. Value of $\bm{y}$ is exactly equal to the undesired signal above the threshold, since
\[
\max\limits_n \left| x(n) - y(n) \right| = \max\limits_n \left| x(n) - x(n) \left( 1 - \frac{\tau}{\left| x(n) \right|} \right) \right| = \tau.
\]
So our task is then to fit $\bm{y}$ as closely as possible.

The signal with reduced PAPR can be calculated as:
\begin{equation}\label{STR2}
\widehat{\bm x} = \bm x - \bm{\Delta x},\\
\end{equation}
where $\bm{\Delta x}$ is the PAPR reduction signal, which we want to be close to $\bm{y}$. Considering a limited bandwidth for $\bm{\Delta x}$ results in
\begin{equation}\label{STR22}
  \bm{\Delta x} = \bm{S}\bm{a},\\
\end{equation}
where $\bm{S} \in \mathbb{C}^{N_{FFT}\times N_{SC}}$ is the matrix, consisting of selected subcarrier $e^{\frac{2\pi \sqrt{-1}mn}{N_{FFT}}}$, $m\in \left[1\dots N_{SC}\right]$ is the selected subcarrier index, $N_{SC}$ is the number of selected subcarriers in the symbol and vector $\bm{a}$ is the amplitudes of these subcarriers. In other words, vector $\bm{a}$ represents amplitudes of the PAPR reduction noise in the frequency domain. Generally, $\bm{a}$ can be computed using LS as follows:
\begin{equation}\label{STR3}
\begin{aligned}
\bm{a} & = \mathop{\arg \min}\limits_{\bm{a} \in \mathbb{C}^{N_{SC}}} {{\left\|{\bm{y}}-{\bm{S}\bm{a}}\right\|}_{2}^2} \\
& = ({{\bm{S}}^H{\bm{S}})^{-1}{\bm{S}}^H{\bm{y}}} \\
& = \frac{1}{N_{FFT}} {{\bm{S}}^H{\bm{y}}}.
\end{aligned}
\end{equation}
A major drawback of Eqs. \eqref{STR1}-\eqref{STR3} is a high computational complexity. Even though matrix product can be calculated using Inverse Fast Fourier Transform (FFT):
\begin{equation}\label{FFT}
{{\bm{S}}^H{\bm{y}} =  \mathcal{F}({\bm{y}})},
\end{equation}
with further taking the achieved FFT outputs according to the reserved subcarrier set, its complexity $\mathcal{O} \left( N_{ITER} N_{ANT} N_{FFT} N_{UP} \log \left( N_{FFT} N_{UP} \right) \right)$ is still too high, where $N_{ITER}$ is the number of PAPR reduction iterations and $N_{UP}$ is the upsampling factor.

\subsection{Sparse STR}

Combining of Eqs. \eqref{STR2}, \eqref{STR22} and \eqref{STR3} results in
\begin{equation}\label{STR4}
\begin{aligned}
\widehat{\bm x} &= \bm x - \frac{1}{N_{FFT}}\bm{L}{\bm{y}}\\
\bm{L}&=\bm{S}{\bm{S}}^H,
\end{aligned}
\end{equation}
where the matrix $\bm{L} \in \mathbb{C}^{N_{FFT}\times N_{FFT}}$ consists of elements 
\begin{equation}\label{STR5}
L_{ij}=\sum_{m=1}^{N_{SC}}e^{\frac{2 \pi \sqrt{-1}m}{N_{FFT}}(i-j)}.
\end{equation}
Now let's analyze the case of only $1$ non-zero element in the vector $\bm{y}$ at the position $n$. In this case, $j=n$ according to Eqs. \eqref{STR4}-\eqref{STR5}, and $i$-th element of PAPR compensation vector $\bm{\Delta x}$ is calculated as
\begin{equation}\label{STR6}
{\Delta x}_i=\frac{y_n}{N_{FFT}}\sum_{m=-N_{SC}/2}^{N_{SC}/2}e^{\frac{2 \pi \sqrt{-1}m}{N_{FFT}}(i-n)}.
\end{equation}
The above equation represents the geometrical progression with parameters ${b_1,q,N_{SC}/2}$. Its sum can be calculated as:
\begin{equation}\label{STR7}
\begin{aligned}
{\Delta x}_i &= \frac{y_n}{N_{FFT}}b_1 \frac{1-q^{N_{SC}/2}}{1-q}\\
&=y_n\frac{N_{SC}}{N_{FFT}}e^{\frac{\pi\sqrt{-1}(-N_{SC}+1)(i-n)}{N_{FFT}}}\frac{1-q^{N_{SC}/2}}{1-q}\\
&=y_n\frac{N_{SC}}{N_{FFT}} \frac{\sin{\frac{\pi N_{SC}(i-n)}{N_{FFT}}}}{\sin{\frac{\pi (i-n)}{N_{FFT}}}}\\
&=\frac{N_{SC}}{N_{FFT}}y_n \operatorname{SINC}(i-n),
\end{aligned}
\end{equation}
where 
\[
\operatorname{SINC}(i-n) = \frac{\sin{\frac{\pi N_{SC}(i-n)}{N_{FFT}}}}{\sin{\frac{\pi (i-n)}{N_{FFT}}}}
\]
is the digital SINC function circularly shifted to the peak position $n$. The SINC can be calculated in advance by taking the IFFT of the rectangle frequency domain signal defined in the range $[-\frac{N_{SC}}{2}...-\frac{N_{SC}}{2}]$.

Suppose having $K$ peaks in $\bm{y}$ at time positions $n_k$. In this case, eq. \eqref{STR7} is transformed to the sum of weighted $\operatorname{SINC}$ functions:
\begin{equation}\label{STR8}
{\Delta x}_i=\frac{N_{SC}}{N_{FFT}} \sum_{k=1}^{K} y(n_k) \operatorname{SINC}(i-n_k).
\end{equation}
The main advantage of the algorithm \eqref{STR8} is a reduced complexity because the matrix product employed in \eqref{STR4} is not required.

Let's calculate the reduction of the peak at $n$-th position. For that, assume orthogonality between $\operatorname{SINC}$ functions. Note that the interference between neighbor SINCs can be neglected if it is much less than the SINC amplitude. The assumption is valid when the delay between peaks is high enough. 


Then we find the PAPR reduction signal value \eqref{STR8} at index $i=n$:
\begin{equation}\label{eq:PAPRscal}
{\Delta x}_i(i=n)=y_n\frac{N_{SC}}{N_{FFT}}.
\end{equation}
Therefore, an initial peak value $y_n$ is reduced to the value of $y_n(1-\frac{N_{SC}}{N_{FFT}})$. As a result, for all peaks we have the same scaling by the factor of $(1-\frac{N_{SC}}{N_{FFT}})$. To have a factor of 1 we can multiply $\bm{y}$ by the factor $\frac{N_{FFT}}{N_{SC}}$. Then it cancels $\frac{N_{SC}}{N_{FFT}}$ factor in \eqref{eq:PAPRscal}.

As a result, \eqref{STR8} becomes
\begin{equation}\label{STR11}
\begin{aligned}
{\Delta x}_i &= \sum_{k=1}^{K} y(n_k) \operatorname{SINC}(i-n_k), \\
i &=1...N_{FFT}.
\end{aligned}
\end{equation}

Thus, to limit the signal amplitude by the value $\tau$ one can use the STR algorithm \eqref{STR1}-\eqref{STR3} with $\bm{y}$ rescale by the factor $\frac{N_{FFT}}{N_{SC}}$ or simply (baseline of the current paper) subtract weighed $\operatorname{SINC}$s using eqs.\eqref{STR1}, \eqref{STR2} and \eqref{STR11}.

\subsection{Divide-and-conquer}\label{div}

The complexity of the algorithm described by \eqref{STR1}, \eqref{STR2} and \eqref{STR11} is mostly defined by the search of $K$ (unknown parameter) time domain positions (delays) of $\operatorname{SINC}$ functions in vector $\bm {|y|}$ because using $y(n_k)$ requires the knowledge of $n_k$. 

As far as the PAPR reduction requires about $4-8$ times upsampling, the bandwidth can be much less than the sampling frequency. In this case, the main lobe of the $\operatorname{SINC}$ function represents not a delta function but can contain several points. Therefore, nearby to a sample $n_k$ with $|y(n_k)|>\tau$, several samples of $\bm y$ can also exceed $\tau$. Thus, the peak selection algorithm should not only find indices $n_k$ with $|y(n_k)|>\tau$, but also find an extremum (maximum) inside each group of such samples. Otherwise, PAPR reduction complexity may grow because it linearly depends on the number of peaks $K$. 


Considering no peak interference, assume the vector $\bm y$ can be divided into $N_{B}>K$ time-domain blocks, where each block contains $\leq 1$ maximums. As a result, \eqref{STR11} becomes
\begin{equation}\label{STR12}
\begin{aligned}
{\Delta x}_i &= \sum_{m=1}^{N_{B}} y(n_m) \operatorname{SINC}(i-n_m), \\
i &=1...N_{FFT}.
\end{aligned}
\end{equation}
where $y(n_m)$ is the element having the maximal absolute value inside $m$-th block, $m$ is the block index. 

Using a few groups $N_{B}$ can reduce the PAPR reduction complexity dramatically, but can also increase the number of $N_{ITER}$ in case any block contains multiple peaks. 
In case a found maximum inside a block doesn't satisfy the extremum condition (e.g. when the maximum is located at the edge of the block), it still can be assumed as a maximum at the cost of negligible increase in PAPR reduction iterations. 

\section{Multiple antennas - Hybrid beamforming}

\begin{figure}[t!]
\centering
\includegraphics[width=0.99\columnwidth]{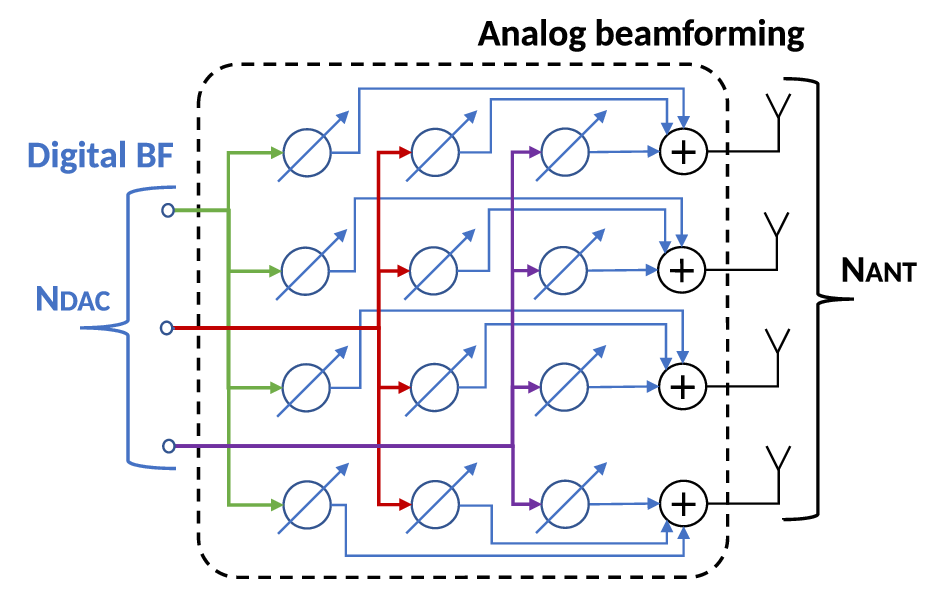}
\caption{Fully-connected HBF.
}
\label{HBF_scheme}
\end{figure}

In ``fully-connected'' HBF, all digital ports are connected to all antennas via phase shifters, as shown in Fig.\ref{HBF_scheme}. This is the generalized case of HBF where $N_{DAC}$ digital streams produce a signal in $N_{ANT}$ antennas as follows:
\begin{equation}\label{HBF}
\bm X = \bm P \bm Z\\,
\end{equation}
where $\bm X \in \mathbb{C}^{N_{ANT} \times N_{FFT}}$ is the time domain multi-antenna signal, $\bm P \in \mathbb{C}^{N_{ANT} \times N_{DAC}}$ is the transformation (analog precoding) matrix, $\bm Z \in \mathbb{C}^{N_{DAC} \times N_{FFT}}$ is the signal produced by digital-to-analog (DAC) converters. Elements of $\bm P$ represent the phase shifter's result $e^{\sqrt{-1}\phi(i_1,i_2)}$, $i_1$ is the DAC index, $i_2$ is the antenna index. 

Firstly, reducing PAPR in the HBF system requires knowledge of the antenna signal $\bm X$ because this signal exists only in the analog domain. It can be calculated using eq. \eqref{HBF} (so-called digital ``twin'' of the antenna signal).

Using eq. \eqref{HBF}, we have
\begin{equation}\label{HBF2}
\begin{aligned}
\widehat{\bm X}&=\bm P \widehat{\bm Z}\\
\bm X-\bm {\Delta X} &=\bm P (\bm Z - \bm {\Delta Z}), \\
\bm {\Delta X} &=\bm P \bm {\Delta Z},
\end{aligned}
\end{equation}

Generally, the PAPR reduction problem in multi-antenna systems is formulated as
\begin{equation}\label{HBF22}
\bm {\Delta Z} = \mathrm{argmin} \max{|\widehat{\bm X}|},
\end{equation}
i.e. find the DAC signal $\bm {\Delta Z}$ minimizing the maximal absolute value among all elements of $\widehat{\bm X}$. 

As an approximation of eq. \eqref{HBF22} solution, we firstly calculate the PAPR reduction signal $\bm{\Delta x}$ for each antenna using \eqref{STR12} and then construct a full $\bm{\Delta X}$ matrix. Having $\bm{\Delta X}$ and $\bm{P}$ allows getting the $\bm {\Delta Z}$ estimate.

Let's analyze the precoding matrix $\bm P$. This matrix consists of normalized orthogonal precoding vectors (beams) that divide users into clusters in the spatial domain. Moreover, beam elements are $e^{\sqrt{-1}\phi(i_1,i_2)}$. Thus, without loss of generality, assume $\bm P$ is the submatrix of the FFT matrix. Generally, $\bm {\Delta Z}$ can be computed using LS as follows:
\begin{equation}\label{HBF4}
\begin{aligned}
\bm {\Delta Z} & = \mathrm{argmin}{{\left\|{\bm {\Delta X}}-{\bm{P}\bm {\Delta Z}}\right\|}_{2}^2} \\
&=(\bm P^H \bm P)^{-1}\bm P^H \bm {\Delta X},\\
&=\frac{1}{N_{ANT}}\bm P^H \bm {\Delta X}
\end{aligned}
\end{equation}
According to eq. \eqref{STR11}, peak amplitudes and positions are known. Therefore, instead of straightforward transforming $\bm{\Delta X}$ from the antenna to the DAC domain described in eq. \eqref{HBF4}, one can transform only amplitudes of $\operatorname{SINC}$s with further multiplication of them by shifted $\operatorname{SINC}$s at the final stage to derive $\bm {\Delta Z}$.

Now let's examine eq. \eqref{HBF4} performance. 
\begin{itemize}
\item Peak selection is performed in the antenna space for each antenna independently, but their reduction is implemented in the DAC subspace, i.e. a subspace of much lower size $(N_{DAC}<N_{TX})$. Therefore, we can't generate $\bm{\Delta Z}$ that fully corresponds to $\bm{\Delta X}$. As a result, all antenna peaks in $\bm X$ will be under-reduced because LS algorithm \eqref{HBF4} also avoids the noise at $\bm{\Delta X}$ samples having zero amplitude.
\item Eq. \eqref{HBF4} implements the LS approximation of peak amplitudes instead of minimizing the maximum peak magnitude. As we have already found from Section \ref{sec:model}, the performance of LS-based solution can be far from the performance of solution achieved by convex optimization of eq. \eqref{HBF22}.
\end{itemize}

\subsection{Trainable LS}\label{train}

\subsubsection{ LS1} One way of improving the PAPR reduction algorithm \eqref{HBF4} performance is the $\bm {\Delta Z}$ scaling ($\rm{coef}>1$). It helps to solve the under-reduction at the cost of noise amplitude increase in samples $\bm X$ whose amplitudes were below the threshold $\tau$.  

\subsubsection{ LS2} Another method presumes excluding all zeros from the vector $\bm {\Delta x}$, consisting of $\operatorname{SINC}$ amplitudes. As a result, the modified $\bm {\Delta \tilde x}$ has a lower size. Peak indexes $n_k$ are known for each antenna, thus no sorting is required. For each time domain index $i$ we can calculate the DAC vector  
\begin{equation}\label{HBF5}
\begin{aligned}
\bm {\Delta z}_i &=\frac{\rm{coef}}{N_{ANT}}\bm P_i^H \bm {\Delta \tilde x}_i,\\
i&=1...N_{FFT}.
\end{aligned}
\end{equation}
In practice, $N_{DAC}>K$, therefore, there is no under-reduction problem, i.e. peaks will be reduced down to $\tau$. But the drawback of algorithm \eqref{HBF5} is the same: noise amplitude increase in samples $\bm X$ if amplitudes were below $\tau$. Thus, $\bm {\Delta z}_i$ scaling ($\rm{coef}<1$) can help to find an optimal value.

\begin{figure}[t!]
\centering
\includegraphics[width=0.99\columnwidth]{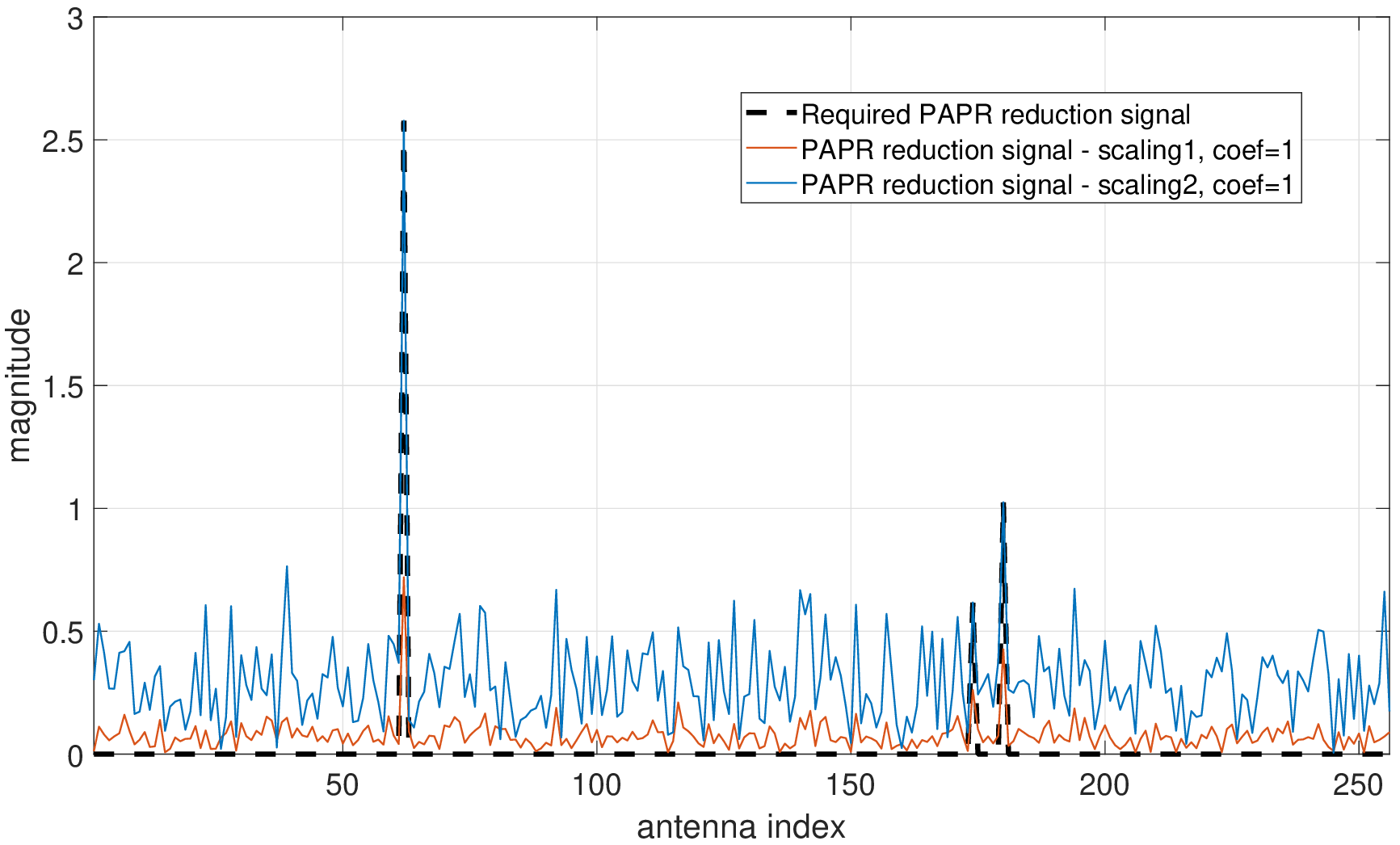}
\caption{PAPR reduction signal in the antenna domain. Generated using \textit{Scaling 1} and \textit{Scaling 2} ($\rm{coef}=1$ for both) and compared to the required one.
}
\label{coef}
\end{figure}

Both \textit{LS1} and \textit{LS2} methods without scaling ($\rm{coef}=1$) are presented in Fig. \ref{coef} 

\subsection{Performance bound}

The problem of finding the optimal PAPR reduction signal lies in considering both a limited bandwidth and a limited digital space. To validate algorithms proposed in the paper, we employed CVX: Matlab-based modeling system for disciplined convex optimization \cite{cvx,Notes}. CVX turns Matlab into a modeling language, allowing constraints and objectives to be specified using standard Matlab expression syntax.

The algorithm \ref{alg} was implemented in CVX and tested in Matlab to find the optimal $\bm {\Delta Z}$ value. The corresponding code is presented in Algorithm \ref{alg}, where $\operatorname{SINC\_mtx} \in \mathbb{C}^ {N_{FFT} \times N_{FFT}}$ is the matrix consisting of circularly shifted copies of the $\operatorname{SINC}$ function. In this paper, this bound was tested in $2$ versions: with $\operatorname{SINC}$ and without it to demonstrate the spectrum limitation impact on the PAPR reduction performance.

\begin{algorithm}[t!]
\caption{Convex optimization in Matlab-CVX}\label{alg}
\begin{algorithmic}
\State \text{cvx\_precision low}
\State \text{cvx\_begin}
\State \text{variable dZ(Ndac,Nfft) complex}
\State \quad \text{minimize(norm(reshape(X - P*dZ*SINC\_mtx,1,[]),'inf'))}
\State \quad \text{subject to norm(dZ,'fro') $<=$ max\_power}
\State \text{cvx\_end} 
\end{algorithmic}
\end{algorithm}

\section{Step-by-step algorithm and its parameters}\label{sec-algo}

Step-by-step implementation of the PAPR reduction algorithm proposed in this paper is shown in Algorithm \ref{steps}. ``LS2'' version of trainable LS proposed in Sec. \ref{train} was employed because of low variation in $\operatorname{coef}$ value when training in various scenarios and lower complexity. Upsampling was excluded, i.e. $N_{UP}=1$, because $N_{FFT}\gg N_{SC}$.

\begin{algorithm}[t!]
\caption{PAPR reduction in HBF}\label{steps}
\begin{algorithmic}
\State \text{Calculate digital twin - eq.~\eqref{HBF}}
\For{$1$ to $N_{ITER}$}
\State \text{Do thresholding for each antenna to find peaks  - eq.~\eqref{STR1};}
\State \text{Divide each antenna signal into $N_B$ blocks - Sec.\ref{div};}
\State \text{Find the maximal abs value in each block;}
\State \text{Find the PAPR reduction signal for antennas - eq.~\eqref{STR12};} 
\State \text{Implement PAPR reduction for each antenna - eq.~\eqref{STR2};}
\State \text{Accumulate PAPR reduction signal;}
\EndFor
\State \text{Construct matrix $\bm {\Delta X}$ using $\bm {\Delta x}_i$}
\State \text{Find PAPR reducing signal in the DAC domain - eq.~\eqref{HBF5}}
\State \text{Calculate the desired DAC signal $\widehat{\bm Z}$}
\end{algorithmic}
\end{algorithm}

Algorithm parameters and simulation parameters are presented in Table \ref{table}. 

\section{Computational complexity}\label{param}

\begin{table}[t!]
    \renewcommand{\arraystretch}{1.05}    
        \begin{tabular}{|p{2.0cm}|p{1.5cm}||p{2.0cm}|p{1.5cm}|}\hline
            \textbf{Parameter}    & \textbf{Value}  &  \textbf{Parameter}& \textbf{Value}   \\ \hline
            \textbf{$N_{FFT}$} &  1024     &  \textbf{$N_{ITER}$} & 1...3 \\ 
            \textbf{$N_{SC}$}  &  240      &  \textbf{$N_{B}$} & 32 \\ 
            \textbf{$N_{UP}$}  & 1         &  \textbf{$N_{OFDM}$} & 120 \\ 
            \textbf{$N_{ANT}$} &  256      &  $\operatorname{Modulation}$ & $QAM16$ \\ 
            \textbf{$N_{DAC}$} & 64        &  $EVM$ & $13.5\%$ \\ 
            \textbf{$N_{LPF}$} & 15        &  $\operatorname{coef}$, $\tau$ & trainable \\ 
            \hline
        \end{tabular}
        \caption{Simulation parameters.}
        \label{table}
\end{table}

Let's discuss the complexity of the PAPR reduction in HBF.

The complexity of the digital twin calculation is defined by calculating the digital twin of antenna domain signal \eqref{HBF}. As this operation can be done using FFT, the complexity is $\mathcal{O} \left( N_{FFT} N_{ANT} \log N_{ANT} \right)$. It can't be avoided or minimized.

Then for each antenna, the upsampling is done with the factor of $N_{UP}$. Complexity is $\mathcal{O} (N_{FFT}N_{ANT}N_{UP}N_{LPF})$, where $N_{LPF}$ is the order of low-pass filter utilized for upsampling. After that, the complexity of absolute values calculation at the antenna domain is $\mathcal{O} (N_{FFT}N_{ANT}N_{UP})$. Then each antenna signal is divided into $N_B$ blocks, with further search for only one maximum element in each block. Thus, the complexity of the search for the maximal values is $\mathcal{O} (N_{FFT}N_{ANT}N_{UP})$. Then for each block, we have to multiply $\operatorname{SINC}$ by an amplitude and subtract from $\bm X$ for each antenna in several iterations. It takes $\mathcal{O} (N_{ITER}N_{FFT}N_{ANT}N_{UP}N_{B})$. 

Finally, after finding the PAPR reduction signal \eqref{STR11} for each antenna, we transform it to the DAC domain, scale by $\operatorname{coef}$, downsample, and update $\bm Z$. Transformation to the DAC domain takes $\mathcal{O} (N_{FFT} N_{ANT} \log N_{ANT})$, and other costs are negligible.

Thus, the baseline PAPR reduction complexity is $\mathcal{O} (N_{FFT} N_{ANT} \log N_{ANT})$ (digital twin calculation) because it can't be lowered. And the most critical part of our procedure is $\mathcal{O} (N_{ITER}N_{FFT}N_{ANT}N_{UP}N_{B})$. The critical complexity can be less than the baseline if $N_{ITER}N_{B}N_{UP} < \log N_{ANT}$. For us, this condition is not satisfied as $2*32*1>10$. 

To reduce complexity, one can apply windowing to $\operatorname{SINC}$ when multiplying it by an amplitude. This would reduce the most critical part from $N_{FFT}$ done to, e.g. $N_{FFT}/8$ or $N_{FFT}/16$ at the cost of negligible spectrum extension. Moreover, the maximum number of peaks exceeding $\tau$ in amplitude that can be processed in a single iteration is $N_{B}=32$ according to the blockwise structure of the algorithm. This value exceeds $K$ (N-of-peaks, eq. \eqref{STR11}) in practice, i.e. some blocks don't have peaks. This would further reduce the complexity from $N_{B}$ to, e.g. $N_{B}/2$, but it depends on the maximal EVM value employed in the system (if it is high, $\tau$ can be low, and a lot of peaks will be above it).

\section{Simulation results}\label{sec-sim}

\begin{figure}[t!]
\centering
\includegraphics[width=0.99\columnwidth]{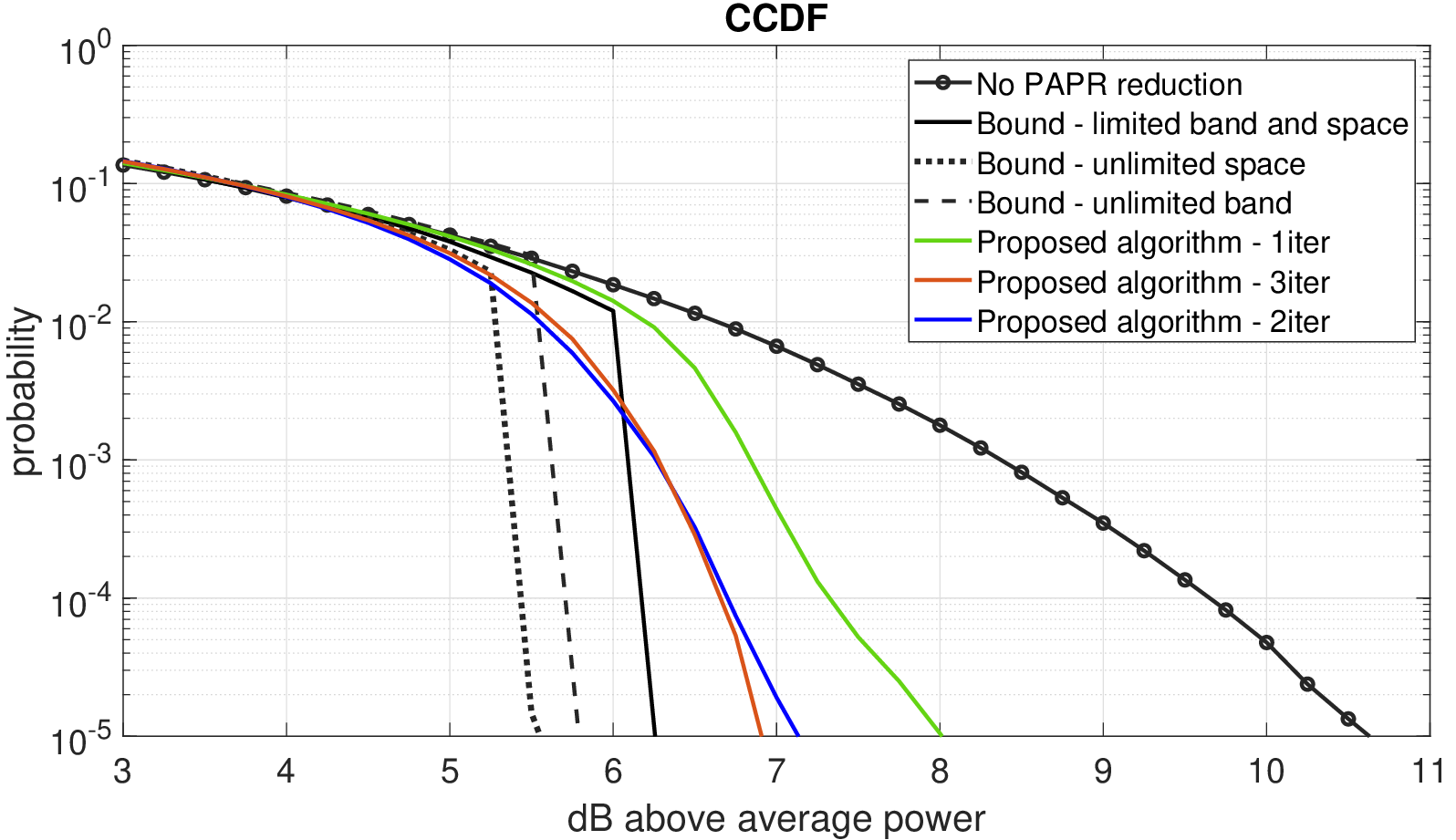}
\caption{CCDF for algorithms and bounds.
}
\label{CCDF}
\end{figure}

In simulations, we tested $1,2$ and $3$ iterations of PAPR reduction. For each $N_{ITER}$ value, parameters $\bm \tau$ and  $\operatorname{coef}$ were trained together by GA to achieve the lowest PAPR at $CCDF=10^{-4}$. The calculated CCDF is presented in Fig. \ref{CCDF}. As we can see, $N_{ITER}=3$ has almost the same PAPR as $N_{ITER}=2$ at $CCDF=10^{-4}$. Therefore, $2$ iterations is enough for PAPR reduction, and the vector of optimal parameters is 
\[
[\operatorname{coef}, \tilde{\tau_1}, \tilde{\tau_2}]=[0.85, 1.76, 1.68], 
\]
where $\tilde{\tau}$ is the normilized value ${\tau}$. The real value of $\tau$ in eq. \ref{STR1} is calculated as 
\[
\tau = \tilde{\tau} \left\| \bm x \right\|,
\] 
where $||\bm x||$ is the $L^2$-norm of $\bm x$. As one can see, $\tilde{\tau}$ slightly depends on the iteration index. 

\begin{figure}[t!]
\centering
\includegraphics[width=0.99\columnwidth]{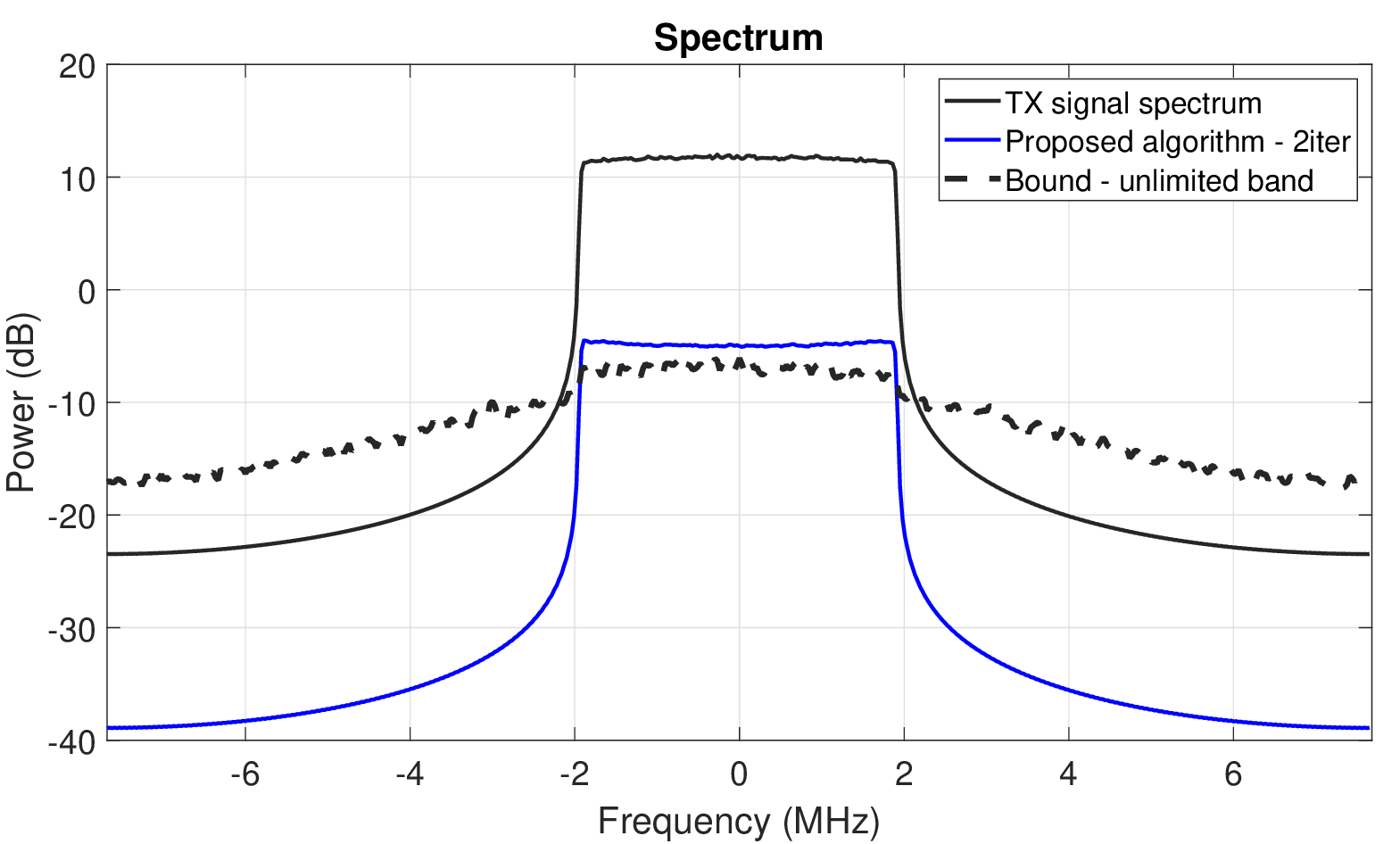}
\caption{Spectrum for the TX signal and PAPR reduction signals (realistic and unlimited band bound).
}
\label{spectrum}
\end{figure}

To justify the algorithm performance we also calculated $3$ practical bounds using the convex optimization described in Algorithm \ref{alg}. Results are presented in Fig. \ref{CCDF}, where \textbf{bound - limited band and space} is our case as the bandwidth is limited and using matrix $\bm P$ limits the space. At $CCDF=10^{-4}$ the PAPR of the proposed method is about $0.5dB$ higher than the bound value (our performance loss). \textbf{Bound - unlimited space} corresponds to PAPR being reduced at each antenna independently, but the bandwidth is limited. This corresponds to a common fully digital single antenna PAPR reduction. Comparing these two bounds, we conclude that the bound shifted by $0.8dB$. Therefore, using $N_{DAC}$ digital outputs to control $N_{ANT}$ (i.e. HBF scheme) increased the PAPR by at least $0.8dB$ compared to the fully digital beamforming. And, finally, \textbf{bound - unlimited band} tells us that PAPR reduction becomes better as more frequency resources are utilized. Out of interest, the spectrum of the \textbf{bound - unlimited band} is presented in Fig.\ref{spectrum}. 

Let us note, that the power of noise caused by PAPR reduction is the same for all techniques and corresponds to the EVM presented in Table \ref{param}.  

\section{Conclusion}

The main challenge of PAPR reduction in HBR transmitters is the simultaneous consideration of both a limited band and a limited subspace for signal generation. Here we divided the problem correspondingly into two parts. Initially, the limited band property was employed and then we utilized a genetic algorithm to find the scaling coefficient for an LS-based solution to fit it into a limited subspace. Simulation results demonstrate the proposed algorithm for PAPR reduction in HBF has only $0.5dB$ loss compared to the bound achieved by convex optimization. The algorithm requires only $2$ iterations and its complexity is not much higher than the best possible, and there is a window to further reduce it.

\section{Acknowledgment}

The authors acknowledge the use of computational cluster Zhores \cite{Zhores2019} for obtaining the results presented in this paper.


\bibliography{PAPR_fit}

\begin{thebibliography}{10}
\providecommand{\url}[1]{#1}
\csname url@samestyle\endcsname
\providecommand{\newblock}{\relax}
\providecommand{\bibinfo}[2]{#2}
\providecommand{\BIBentrySTDinterwordspacing}{\spaceskip=0pt\relax}
\providecommand{\BIBentryALTinterwordstretchfactor}{4}
\providecommand{\BIBentryALTinterwordspacing}{\spaceskip=\fontdimen2\font plus
\BIBentryALTinterwordstretchfactor\fontdimen3\font minus \fontdimen4\font\relax}
\providecommand{\BIBforeignlanguage}[2]{{%
\expandafter\ifx\csname l@#1\endcsname\relax
\typeout{** WARNING: IEEEtran.bst: No hyphenation pattern has been}%
\typeout{** loaded for the language `#1'. Using the pattern for}%
\typeout{** the default language instead.}%
\else
\language=\csname l@#1\endcsname
\fi
#2}}
\providecommand{\BIBdecl}{\relax}
\BIBdecl

\bibitem{5G}
{Shafi, Mansoor and Molisch, Andreas F. and Smith, Peter J. and Haustein, Thomas and Zhu, Peiying and De Silva, Prasan and Tufvesson, Fredrik and Benjebbour, Anass and Wunder, Gerhard}, ``{5G: A Tutorial Overview of Standards, Trials, Challenges, Deployment, and Practice},'' \emph{IEEE Journal on Selected Areas in Communications}, vol.~35, no.~6, pp. 1201--1221, 2017.

\bibitem{hpa}
A.~Ivanov and D.~Lakontsev, ``Adaptable look-up tables for linearizing high power amplifiers,'' in \emph{2017 3rd International Conference on Frontiers of Signal Processing (ICFSP)}, 2017, pp. 96--100.

\bibitem{DAC_ADC_PAPR}
A.~A. Abdulhussein and H.~N. Abdullah, ``{Comparative Study of Peak to Average Power Ratio in OFDM and FBMC Systems},'' in \emph{2021 7th International Conference on Space Science and Communication (IconSpace)}, Nov 2021, pp. 140--145.

\bibitem{Molisch}
A.~F. Molisch, V.~V. Ratnam, S.~Han, Z.~Li, S.~L.~H. Nguyen, L.~Li, and K.~Haneda, ``Hybrid beamforming for massive mimo: A survey,'' \emph{IEEE Communications Magazine}, vol.~55, no.~9, pp. 134--141, 2017.

\bibitem{Lyashev}
R.~Semernya, S.~Xueliang, V.~Lyashev, V.~Revutsky, Z.~Yue, and D.~Lei, ``Application of hbf with adaptive port mapping for leo satellite communication systems,'' in \emph{2021 IEEE 94th Vehicular Technology Conference (VTC2021-Fall)}, 2021, pp. 1--6.

\bibitem{Sohrabi}
F.~Sohrabi and W.~Yu, ``Hybrid digital and analog beamforming design for large-scale antenna arrays,'' \emph{IEEE Journal of Selected Topics in Signal Processing}, vol.~10, no.~3, pp. 501--513, 2016.

\bibitem{yaliny}
A.~Ivanov, V.~Teplyakov, and V.~Kalinin, ``Mobile communication system with a hybrid phased array antenna system,'' in \emph{2015 IEEE East-West Design \& Test Symposium (EWDTS)}, 2015, pp. 1--4.

\bibitem{phy_hbf}
A.~Ivanov, M.~Stoliarenko, A.~Savinov, and S.~Novichkov, ``Physical layer representation in leo satellite with a hybrid multi-beamforming,'' in \emph{2019 15th International Wireless Communications \& Mobile Computing Conference (IWCMC)}, 2019, pp. 140--145.

\bibitem{Ahmed}
I.~Ahmed, H.~Khammari, A.~Shahid, A.~Musa, K.~S. Kim, E.~De~Poorter, and I.~Moerman, ``A survey on hybrid beamforming techniques in 5g: Architecture and system model perspectives,'' \emph{IEEE Communications Surveys \& Tutorials}, vol.~20, no.~4, pp. 3060--3097, 2018.

\bibitem{CF1}
B.~Lee and Y.~Kim, ``\BIBforeignlanguage{English}{{An adaptive clipping and filtering technique for PAPR reduction of OFDM signals}},'' \emph{\BIBforeignlanguage{English}{Circuits, Systems, and Signal Processing}}, vol.~32, no.~3, pp. 1335--1349, Jun. 2013.

\bibitem{CF2}
A.~Chakrapani and V.~Palanisamy, ``{A Novel Clipping and Filtering Algorithm Based on Noise Cancellation for PAPR Reduction in OFDM Systems},'' \emph{Proceedings of the National Academy of Sciences, India Section A: Physical Sciences}, vol.~84, pp. 467--472, 2014.

\bibitem{PW1}
J.~Hou, J.~Ge, and F.~Gong, ``{Tone Reservation Technique Based on Peak-Windowing Residual Noise for PAPR Reduction in OFDM Systems},'' \emph{IEEE Transactions on Vehicular Technology}, vol.~64, no.~11, pp. 5373--5378, 2015.

\bibitem{TR_FISTA}
Y.~Wang, S.~Xie, and Z.~Xie, ``{FISTA-Based PAPR Reduction Method for Tone Reservation’s OFDM System},'' \emph{IEEE Wireless Communications Letters}, vol.~7, no.~3, pp. 300--303, 2018.

\bibitem{STR}
A.~Ivanov and D.~Lakontsev, ``{Selective tone reservation for PAPR reduction in wireless communication systems},'' in \emph{2017 IEEE International Workshop on Signal Processing Systems (SiPS)}, 2017, pp. 1--6.

\bibitem{TR_DL}
L.~Li, C.~Tellambura, and X.~Tang, ``Improved tone reservation method based on deep learning for papr reduction in ofdm system,'' in \emph{2019 11th International Conference on Wireless Communications and Signal Processing (WCSP)}, Oct 2019, pp. 1--6.

\bibitem{TR_DL1}
X.~Wang, N.~Jin, and J.~Wei, ``A model-driven dl algorithm for papr reduction in ofdm system,'' \emph{IEEE Communications Letters}, vol.~25, no.~7, pp. 2270--2274, 2021.

\bibitem{PAPR_MIMO_DL}
R.~Zayani, H.~Shaiek, and D.~Roviras, ``Papr-aware massive mimo-ofdm downlink,'' \emph{IEEE Access}, vol.~7, pp. 25\,474--25\,484, 2019.

\bibitem{UBR}
A.~Ivanov, A.~Volokhatyi, D.~Lakontsev, and D.~Yarotsky, ``{Unused Beam Reservation for PAPR Reduction in Massive MIMO System},'' in \emph{2018 IEEE 87th Vehicular Technology Conference (VTC Spring)}, 2018, pp. 1--5.

\bibitem{IVANOV2}
A.~Kalinov, R.~Bychkov, A.~Ivanov, A.~Osinsky, and D.~Yarotsky, ``{Machine Learning-Assisted PAPR Reduction in Massive MIMO},'' \emph{IEEE Wireless Communications Letters}, vol.~10, no.~3, pp. 537--541, 2021.

\bibitem{krikunov}
S.~Krikunov, R.~Bychkov, A.~Blagodarnyi, and A.~Ivanov, ``Clustering and fitting to reduce papr in multi-user ofdm systems,'' in \emph{2023 25th International Conference on Digital Signal Processing and its Applications (DSPA)}, 2023, pp. 1--6.

\bibitem{cvx}
M.~Grant and S.~Boyd, ``{CVX}: Matlab software for disciplined convex programming, version 2.1,'' \url{http://cvxr.com/cvx}, Mar. 2014.

\bibitem{Notes}
M.~Grant, ``Graph implementations for nonsmooth convex programs,'' in \emph{Recent Advances in Learning and Control}, ser. Lecture Notes in Control and Information Sciences, V.~Blondel, S.~Boyd, and H.~Kimura, Eds.\hskip 1em plus 0.5em minus 0.4em\relax Springer-Verlag Limited, 2008, pp. 95--110, \url{http://stanford.edu/~boyd/graph_dcp.html}.

\bibitem{Zhores2019}
I.~{Zacharov}, R.~{Arslanov}, M.~{Gunin}, D.~{Stefonishin}, A.~{Bykov}, S.~{Pavlov}, O.~{Panarin}, A.~{Maliutin}, S.~{Rykovanov}, and M.~{Fedorov}, ``Zhores - petaflops supercomputer for data-driven modeling, machine learning and artificial intelligence installed in skolkovo institute of science and technology,'' in \emph{Open Engineering}, vol. 9(1), 2019, pp. 512--520.

\end{thebibliography}
\bibliographystyle{IEEEtran}

\end{document}